\documentclass[aps,prl,nofootinbib,preprintnumbers,amsmath,amssymb,latexsym,array,enumerate,letter,twocolumn,superscriptaddress]{revtex4-1}
\usepackage{amssymb}
\usepackage{amsmath}
\usepackage{epsfig}
\usepackage{hyperref}
\usepackage{breakurl}
\usepackage{color}
\makeatletter
\def\simgt{\mathrel{\lower2.5pt\vbox{\lineskip=0pt\baselineskip=0pt
           \hbox{$>$}\hbox{$\sim$}}}}
\def\simlt{\mathrel{\lower2.5pt\vbox{\lineskip=0pt\baselineskip=0pt
           \hbox{$<$}\hbox{$\sim$}}}}
\makeatother

\begin{document}
\title{\boldmath Jet Topology} 

\author{Lingfeng Li}
\email{iaslfli@ust.hk}
\affiliation{Jockey Club Institute for Advanced Study, 
The Hong Kong University of Science and Technology, Hong Kong S.A.R.}

\author{Tao Liu}
\email{taoliu@ust.hk}
\affiliation{Department of Physics, The Hong Kong University of Science and Technology, Hong Kong S.A.R.}

\author{Si-Jun Xu}
\email{sxuaw@connect.ust.hk}
\affiliation{Department of Physics, The Hong Kong University of Science and Technology, Hong Kong S.A.R.}

\begin{abstract}
We introduce persistent Betti numbers to characterize topological structure of jets. These topological invariants measure multiplicity and connectivity of jet branches at a given scale threshold, while their persistence records evolution of each topological feature as this threshold varies. With this knowledge, in particular, we are able to reconstruct branch phylogenetic tree of each jet. These points are demonstrated in the benchmark scenario of light-quark versus gluon jets. This study provides a topological tool to develop jet taggers, and opens a new angle to look into jet physics. 

\end{abstract}

\maketitle
\section{Introduction}

\begin{figure}[h!]
\centering
\includegraphics[width=1 \columnwidth]{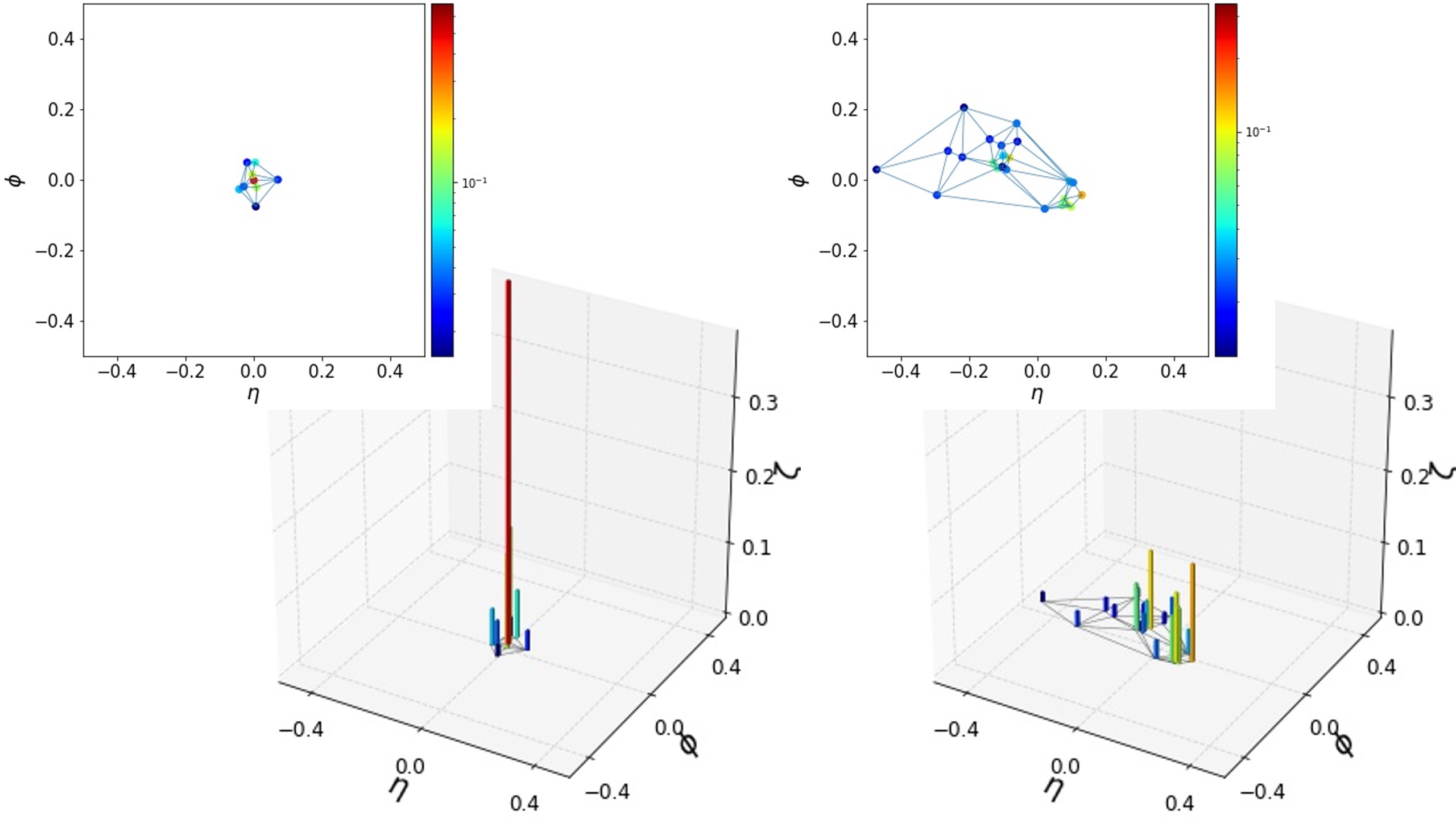}
\caption{Profiles of typical light-quark (left) and gluon (right) jets and their Delaunay triangulations.}
\label{fig:Eflow}
\end{figure}

Jets, either QCD~\cite{Bjorken:1969wi,Drell:1969wb,Hanson:1975fe,DeGrand:1977sy} or boosted-heavy~\cite{Butterworth:2008iy,Ellis:2009su,Feige:2012vc} ones, are initiated by their ancestral particles via perturbative parton shower, with the shower-produced partons being subsequently transferred into hadrons via non-perturbative confinement processes~\cite{CATANI1993187,seymour1994searches,Seymour:1997kj,Wobisch:1998wt}. Though standard clustering provides a way to systematically organize the constituents (mainly hadrons reaching detector) of jets, strong motivations exist to look into their internal structure. In essence, the spray of constituents in each jet are a manifestation of the nature (flavor, QCD charge, four momentum, etc.) of its ancestral particle. Their energy profile inherits from the kinematics of the shower-produced partons. Measuring jet structure thus may decipher the relevant information on its ancestral particle. As one can see from Fig.~\ref{fig:Eflow}, light-quark ($q$) and gluon ($g$) jets demonstrate different features in structure. These features can on the one hand assist tagging jet flavor, and on the other hand, deepen our understanding on jet dynamics. 

In last two decades, one of the most exciting achievements in relation to this is the application of jet substructure techniques for tagging boosted heavy jets~\cite{Butterworth:2008iy,Ellis:2009su,Feige:2012vc}. By inspecting kinematics of their subjets, one could distinguish boosted heavy jets from their QCD backgrounds with an upgraded accuracy. Subjets can also be applied as a probe to jet sturcture. One famous example is jet N-subjettiness~\cite{Thaler:2010tr}, where the jet structure is probed by measuring the degree to which the jet radiations and the candidate subjet directions are collimated. Relatively, the tagging of light QCD jets is more challenging, because of the fineness of their internal structure. Given its importance for, e.g., measuring vector boson production with jets~\cite{Khachatryan:2014dea} and Higgs production via vector boson fusion~\cite{Khachatryan:2015bnx}, various methods have been proposed for tagging light QCD jets: track multiplicity, jet angularities, N-subjettiness, two(multi)-point correlators, etc.~\cite{Ellis:2010rwa,Gallicchio:2012ez,CMS:2013kfa,Larkoski:2014pca,Bhattacherjee:2015psa,Davighi:2017hok,Komiske:2017aww,Metodiev:2018ftz}. Recently, with the deep-neural-network (DNN) techniques, more progresses were achieved for tagging both (for a review, see~\cite{Kasieczka:2019dbj,Larkoski:2017jix}).

Different from these impressive efforts, in this letter we will be dedicated to addressing topological aspects of jet structure. Explicitly, we will introduce persistent Betti numbers~\cite{edelsbrunner2000topological,zomorodian2005computing} to characterize topological structure of jets. These topological invariants measure multiplicity and connectivity of jet branches at a given scale threshold, while their persistence records evolution of each topological feature as this threshold varies. Here {\it ``jet branch'' is not a generalization of the concept of subjet}, despite correlation could exist between them in some contexts. Just like a branch of tree, which is disconnected from its other branches above some threshold of height, the jet branch does so above some threshold of scale ($\zeta$). Its definition is topology-based and does not depend on any ``distance'' measure, the corner stone of jet/subjet clustering, of the jet constituents directly. This reveals a great advantage of topological variants as a probe. That is, they allow us to look into not just the internal kinematics of ``fat'' boosted-heavy jets, but also the fine structure of relatively ``narrow'' light-QCD jets, with a considerable resolution.

\section{Persistent Homology and Jets}

The topological features of jet structure can be captured using a connected planar graph, where the jet constituents are viewed as sparse samplings of jet profile and define the vertices of this graph. This idea in a general context has been well established in computational topology~\cite{correa2011towards,martinetz1993competitive,liu2016visualizing}. In this work we will take a Delaunay triangulation (DT), $i.e.$, the dual graph of Voronoi Diagram~\cite{delaunay1934} (the cell of Voronoi diagram in jet physics has a natural explanation as the passive catchment area for each jet constituent, in the $k_t$ jet clustering~\cite{Cacciari:2008gn}). The method of DT has proved to be able to fully preserve the topology of a given topological object~\cite{martinetz1994topology}, for homogeneous sampling, and hence has been widely accepted~\cite{martinetz1994topology,edelsbrunner1994triangulating,fritzke1995growing,bernardini1997sampling}.
With this process, the jet profile is encoded as a continuous function at the $\eta -\phi$ plane, with its topology being well-represented.

Physically, the multiplicity and connectivity of jet branches evolve as the scale threshold $\zeta$ varies. One natural analogue is again the branches of a tree. As the threshold of height decreases from above the tree, one would expect the number of its connected branches above this threshold increases from zero to many and then decreases to one. To manifest this, we define the superlevel and sublevel sets of $\zeta$ for the jet DT graph $G_{\rm ref}$ as 
\begin{eqnarray} 
      G(\zeta)  &=&  G_{\rm ref} \{  {\rm jet \ constituents \ } (i=1, 2, ...)   \mid  \zeta_i \geq \zeta \}  \ ,  \nonumber \\
 \bar G(\zeta)  &=&  G_{\rm ref}  \{ {\rm jet \ constituents \ } (i=1, 2, ...)  \mid  \zeta_i < \zeta \}  \ . 
\end{eqnarray} 
Here $i$ runs over all jet constituents and $\zeta_i$ is the scale value of the $i$-th one. For a given $\zeta$, these two subgraphs of $G_{\rm ref}$ are complementary to each other. The topological features of jet branches above and below $\zeta$ are then encoded as Betti numbers of $G(\zeta)$ and $\bar G(\zeta)$, respectively.  

Betti number $\beta_i$~\cite{Betti1870} is the rank of the $i$-th Homology group in algebraic topology. The latter characterizes the topology of a space based on the relationship between the cycles and its boundaries. Betti numbers are related to the Euler characteristic $\chi$ via the Euler-Poincar$\text{\'{e}}$ formula 
\begin{equation}
\chi = \sum_{p=0}^d (-1)^p \beta_p \ .
\label{eq:poincare}
\end{equation} 
Here $d$ is the dimension of the topological space, and $\beta_0$, $\beta_1$ and $\beta_2$ count the numbers of the connected components, holes and voids of this space, respectively. For the two-dimensional $\eta-\phi$ plane, this formula is reduced to $\chi=\beta_0-\beta_1$, because of $\beta_2 \equiv 0$. We will use $\beta_0$ and $\beta_1$ to measure the topological features of jets. The relevant topological invariants for $G(\zeta)$ and $\bar{G}(\zeta)$ are summarized in Table~\ref{tab:meaning}. Notably, $\bar{\beta}_0(\zeta)$ receives two contributions. One is the holes of $G(\zeta)$, and another one is the connected components of $\bar{G}(\zeta)$ at the boundary of $G_{\rm ref}$. The story is similar for $\beta_0(\zeta)$. With a test, we observe a strong cross-correlation between $\beta_{0,1}(\zeta)$ and $\bar{\beta}_{1,0}(\zeta)$. So we will present $\beta_{0,1}(\zeta)$ in this letter and use $\bar{\beta}_{1,0}(\zeta)$ to assist the calculations only. 

\begin{table}[h!]
\centering
\begin{footnotesize}
\begin{tabular}{ccc}
\hline
$\chi(\zeta)$, $\bar{\chi}(\zeta)$& Euler characteristics of $G(\zeta)$ or $\bar{G}(\zeta)$ &  \\
$\beta_0(\zeta)$, $\bar{\beta}_0(\zeta)$ & Number of connected components in $G(\zeta)$ or $\bar{G}(\zeta)$ & \\
$\beta_1(\zeta)$, $\bar{\beta}_1(\zeta)$ & Number of holes in $G(\zeta)$ or $\bar{G}(\zeta)$ & \\
$\beta_2(\zeta)$, $\bar{\beta}_2(\zeta)$ & Zero for a 2D graph embedded in a plane & \\
$\bar{\beta}_0(\zeta)-\beta_1(\zeta)$ & Number of $\bar{G}(\zeta)$'s connected components at $B(G_{\rm ref})$ &\\
$\beta_0(\zeta)-\bar{\beta}_1(\zeta)$ & Number of $G(\zeta)$'s connected components at $B(G_{\rm ref})$ &\\
\hline
\end{tabular}
\end{footnotesize}
\caption{Topological invariants defined for $G(\zeta)$ and $\bar{G}(\zeta)$. $B(G_{\rm ref})$ is the boundary of the graph $G_{\rm ref}$. }
\label{tab:meaning}
\end{table}

Morse theory~\cite{morse1925relations} plays another important role in this study. It states in our context that the topology of $G(\zeta)$ changes only if $\zeta$ passes some vertex of $G_{\rm ref}$ or jet constituent. This allows the birth, growth (e.g., energy accretion of a branch due to the joining of new leaves or other  branches) and death of each topological feature to be persistently recorded by evaluating the impact of every $G_{\rm ref}$ vertex passed by $\zeta$, as $\zeta$ varies, and hence makes sense of its evolution. Particularly, with the persistent knowledge of $\beta_0$, we will be able to build the branch phylogenetic tree for each jet. 

To demonstrate these points, below we will study one benchmark scenario of light-quark versus gluon jets. The sample jets are generated from the $pp\to Z+q/g$ events with $Z\to \nu\nu$. They are showered with Pythia8~\cite{Sjostrand:2007gs} and clustered with anti-$k_T$ algorithm~\cite{Cacciari:2008gp} ($\Delta R=0.6$), unless otherwise specified. The $q(=u,d,s)$ and $g$ jets are then equally selected from five 50GeV-wide bins, with the jet $p_T$ sequentially ranging from $100{\rm GeV}$ to $350{\rm GeV}$. Additionally, we define $\zeta = \frac{p_T}{p_T^{\rm Jet}}$ as a normalized scale for the convenience of discussions (note, $\zeta$ is not necessary to be based on $p_T$/energy; we will not go to other possibilities such as angular scale in this work). To ensure these topological invariants to be infrared- and collinear-safe technically, we graph the jets by vetoing their soft constituents with $\zeta_i <  10^{-2}$ and merging the collinear constituents with $\Delta R_{ij} < 0.01$. We will not take into any detector effects, given that this is a conceptual study. At last, instead of using the usual persistence length $\zeta_b - \zeta_d$ (i.e., the difference between the birth and death moments), we introduce relative persistence length, i.e., $\zeta_b/\zeta_d$, to measure the (relative) lifetime of a given topological feature.

\section{Light-Quark Vs. Gluon Jets}

\begin{figure}[h!]
\centering
\includegraphics[width=1.0 \columnwidth]{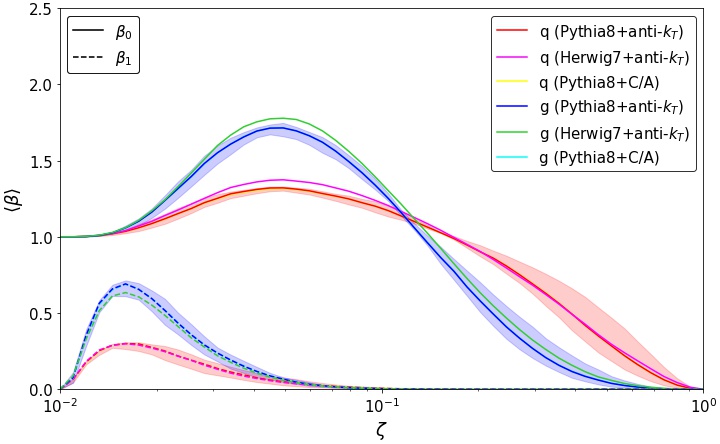}
\caption{Averaged Betti numbers of the super-level sets of $\zeta$, for the light-quark and gluon jets ($100{\rm GeV} < p_T < 350{\rm GeV}$).}
\label{fig:betti_qg}
\end{figure}

The averaged Betti numbers of the superlevel sets of $\zeta$, for the $q$ and $g$ jets, are shown in Fig.~\ref{fig:betti_qg}. The red and blue curves of  $\langle \beta_0 \rangle$ and $\langle \beta_1 \rangle$ in this figure are defined by the jets from all of the five bins, while the bands hosting them are expanded by the five bin-based  averages. The narrowness of these bands indicates that these topological features are not very sensitive to $p_T$ for the jets with $p_T > 100$GeV. As jet $p_T$ drops below 100GeV, the $\langle \beta_0 \rangle$ and $\langle \beta_1 \rangle$ peaks will take a quickening overall downward shift, due to the weakening of parton shower.

Fig.~\ref{fig:betti_qg} demonstrates a series of features distinguishable between the $q$ and $g$ jets. At $\zeta =1$, all jets have $\beta_0 = \beta_1 \equiv 0$ and hence $\langle \beta_0 \rangle=\langle \beta_1 \rangle \equiv  0$. As $\zeta$ decreases, some branch ends become above this threshold and $\langle \beta_0 \rangle$ develops a non-zero value. The $q$-jet $\langle \beta_0 \rangle$ rises up earlier than the $g$-jet one, and then enters a stage of slow evolution. In contrast, the $g$-jet $\langle \beta_0 \rangle$ rises up later, but it grows a peak higher than that of the $q$-jet $\langle \beta_0 \rangle$. This indicates that: the $g$ jets develop more branches, while the primary branch of the $q$ jets tends to be harder. This is consistent with the radiation profiles of the two typical $q$ and $g$ jets shown in Fig.~\ref{fig:Eflow}. In physics gluons carry a larger color charge, which necessarily results in a stronger shower for the ancestral gluons, and correspondingly, a smaller energy share for the shower-produced partons. As $\zeta$ keeps decreasing, more and more soft jet constituents become above this threshold. Some of them serve as a connector, yielding a decrease of $\langle \beta_0 \rangle$ or an increase of $\langle \beta_1 \rangle$. The $g$ jets show a bigger chance to form the holes, compared to the $q$ jets. As $\zeta$ goes down to $10^{-2}$, $\langle \beta_0 \rangle$ evolves to one, as all non-primary branches have been connected to the primary one either directly or indirectly, and $\langle \beta_1 \rangle$ to zero, as all holes filled by some softer constituents. 

To test the robustness of these topological observables against the parton-shower models and jet clustering algorithms, we also plot the $\langle \beta_0 \rangle$ and $\langle \beta_1 \rangle$ curves based on the Herwig7~\cite{Bellm:2015jjp} + anti-$k_T$ and Pythia8 + Cambridge/Aachen (C/A)~\cite{Wobisch:1998wt,Dokshitzer:1997in} combinations in Fig.~\ref{fig:betti_qg}. Different from Pythia8~\cite{Sjostrand:2007gs} which takes $p_T$-ordered dipole showering, Herwig7 is angular-ordered. As for jet clustering, the C/A algorithm does not give a priority to hard jet constituents, unlike anti-$k_T$. One can see from this figure that Herwig7 causes a shift of $\sim 10\%$ level at the peak relative to the Pythia8 + anti-$k_T$ curves, while the C/A yields a tiny impact of percent level only.

\begin{figure}[h!]
\centering
\includegraphics[width=1.0 \columnwidth]{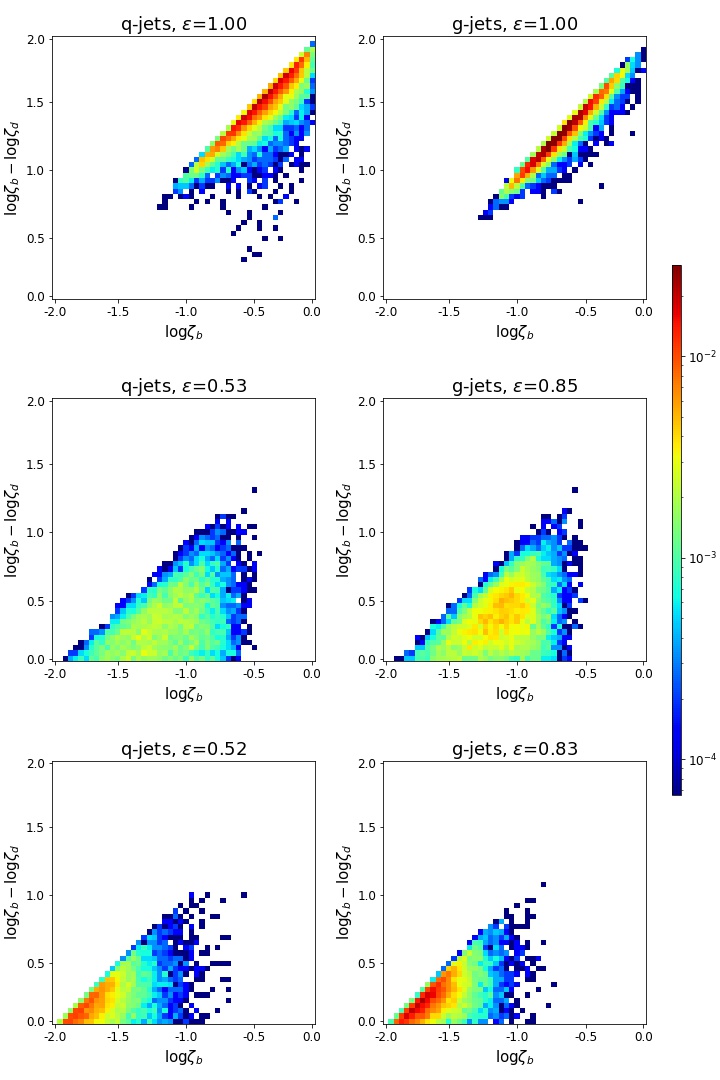}
\caption{Persistence diagrams of the first (upper), second (middle) $\beta_0$ features and the first (bottom) $\beta_1$ feature, for the light-quark (left) and gluon (right) jets. These features are sorted by $\zeta_b/\zeta_d$. $\varepsilon$ is the fraction of the jets in each sample which grow the relevant feature. The color bar represents the normalized density of the scattering points.}
\label{fig:per_qg}
\end{figure}

In Fig.~\ref{fig:per_qg}, we present the persistence diagrams of the first, second $\beta_0$ features and the first $\beta_1$ feature, for the $q$ and $g$ jets. Here the ``death'' moment of the first $\beta_0$ feature (i.e., the primary branch) is defined as the threshold where the last leave/branch is connected in (note, this does not imply a ``real'' death of the first $\beta_0$ feature, but introduces this quantity as an observable), different from the others. From this figure, one can easily see that the first $\beta_0$ feature of the $q$ jets tends to be born earlier than that of the $g$ jets. This is consistent with our discussions on Fig.~\ref{fig:betti_qg}. But, being not explicitly shown in Fig.~\ref{fig:betti_qg}, about one half of the $q$ jets fail to grow the second $\beta_0$ feature (similar for the first $\beta_1$ feature), in comparison to $\sim 15\%$ of the $g$ jets. More than that, these two topological features, if being developed, tend to have a longer lifetime for the $g$ jets than the $q$ jets. 

\begin{figure}[h!]
\centering
\includegraphics[width=1.0 \columnwidth]{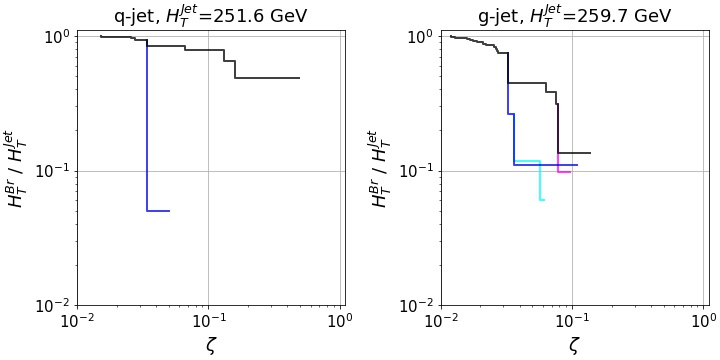}
\caption{Branch phylogenetic trees for the two typical light-quark (left) and gluon (right) jets shown in Fig.~\ref{fig:Eflow}.}
\label{fig:phylo}
\end{figure}

With the persistent knowledge of $\beta_0$, we are able to reconstruct the branch phylogenetic tree for each jet. We show the ones in Fig.~\ref{fig:phylo} for the two typical $q$ and $g$ jets in Fig.~\ref{fig:Eflow}.  This figure is reminiscent of the phylogenetic trees extensively used in epidemiology (see, e.g., the virus phylogenetic trees of GISAID~\cite{gisaid}). As is expected, the $q$-jet tree has fewer branches than the $g$-jet tree, while its primary branch is ``higher'' than the $g$-jet one. For this $q$ jet, all leaves grow on the primary branch, making it very strong (i.e., have a big $H_T^{\rm Br} / H_T^{\rm Jet}$ value; here $H_T^{\rm Br}$ and $H_T^{\rm Jet}$ denote the $p_T$ scalar sum of the branch and the jet, respectively) before another branch is connected in. As a comparison, the phylogenetic structure of this $g$ jet is richer. The energy allocation is also more democratic for the growth of its branches. Eventually, this $g$ jet develops two holes as $\zeta$ decreases, in comparison to zero of the given $q$ jet. The more robust hole is surrounded by its blue-cyan branches and leaves, which can be easily figured out from Fig.~\ref{fig:Eflow}.

\begin{figure}[h!]
\centering
\includegraphics[width=1.0 \columnwidth]{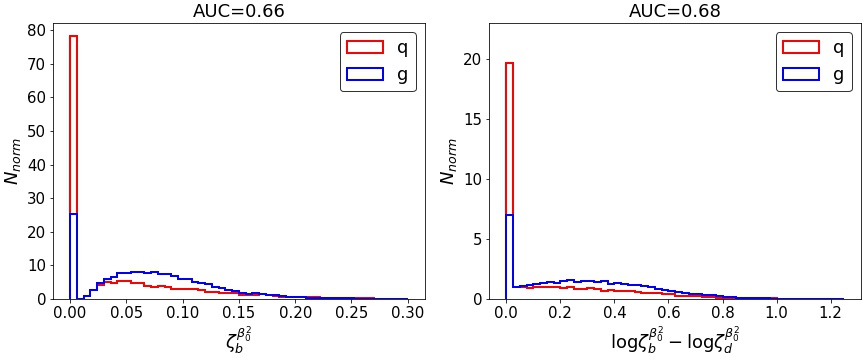}
\caption{Distributions of $\zeta_b^{\beta_0^2}$ (left) and $\log \zeta_b^{\beta_0^2} / \zeta_d^{\beta_0^2}$ (right), for the light-quark and gluon jets. For the jets without the second $\beta_0$ feature, they are counted by the zero bin.}
\label{fig:pro}
\end{figure}

To glance at the potential power of these topological observables to distinguish between the $q$ and $g$ jets, we project their scattering-point distributions of the second $\beta_0$ feature (denoted as $\beta_0^2$) (see the middle panels of Fig.~\ref{fig:per_qg}) to the $\zeta_b^{\beta_0^2}$ and $\log \zeta_b^{\beta_0^2} / \zeta_d^{\beta_0^2}$ axes. Please note that this never means that the other topological features are not important. The projection histograms are presented in Fig.~\ref{fig:pro}, which generate an AUC of 0.66 for $\zeta_b^{\beta_0^2}$ and 0.68 for $\log\zeta_b^{\beta_0^2}/\zeta_d^{\beta_0^2}$. This outcome is encouraging, considering that the $q$ and $g$ jets are simulated within a wide range of $p_T$: $100{\rm GeV} < p_T < 350{\rm GeV}$, and $\zeta_b^{\beta_0^2}$ and $\log \zeta_b^{\beta_0^2}$ are just two of the observables in this topological scheme. As a comparison, the AUCs of the DNN-based $q$-$g$ jet classifiers vary from 0.81 to 0.91~\cite{Komiske:2016rsd,Cheng:2017rdo,Kasieczka:2018lwf,Komiske:2018cqr,Qu:2019gqs,Larkoski:2019nwj}, for different test samples and methods.

\section{Summary and Outlook}

In this letter, we demonstrated how the tool of persistent homology be applied to studying jet physics, using the benchmark scenario of light-quark versus gluon jets. Following this effort, we can immediately see several important directions for next-step explorations.

First of all, optimize this study and build topological $q$- and $g$-jet classifiers. To develop such jet classifiers, we need to properly synergize the observables in this topological scheme (including the information carried by the branch phylogenetic trees), and maybe some complementary others known to us before. This could be achieved by using the DNN techniques. Also, the analysis taken in this study has not been optimized yet. In the framework of Voronoi diagram,  Delaunay triangulation taken in this study is equivalent to defining an energy density in the unit of energy per cell for the jets. This profile does not depend on the cell area directly. Alternatively, one can define an energy density per based on the energy of each jet constituent and the area of its host cell. We do not know yet which one represents better the topological features of jet structure. Additionally, one advantage of persistent topology is that the persistence length can be applied to strengthening the robustness of the relevant topological observables against topological noise. This point has been extensively appreciated by people working on topological computation~\cite{zomorodian2005computing,CARR200375}. In our context, we can apply, e.g., $\log\zeta_b/\zeta_d$,  as a trimming tool to remove the jet branches with a short lifetime. This will allow us to focus on the more robust topological features in each jet, and may benefit to the classifier construction by suppressing topological contaminations from environment (pileups, underlying events, detector noise, etc.).   

Secondly, develop topological taggers for boosted heavy jets. In this work we overwhelmingly focused on light QCD jets. But, it is a natural thinking to extend this study to boosted heavy jets such as $W^\pm$, $Z$, Higgs and top jets, given their important role in searching for new physics at Large Hadron Collider and even future hadron collider.  

Thirdly, develop topological probes to jet dynamics at perturbative level. As its heritage after non-perturbative QCD confining, jet radiation profile lives in a higher-dimensional phase space. Its topological features thus may shed light on hard process and parton splitting. For example, the $g$ jet demonstrated in Fig.~\ref{fig:Eflow} and Fig.~\ref{fig:phylo}, which develops two groups of branches (black-magenta and blue-cyan ones) and also two holes as $\zeta$ decreases, actually originates from a splitting of its ancestral gluon into two roughly-symmetric gluon branches. The tool of persistent homology might provide a handle to probabilistically unfold the relevant physics  (for probabilistic approaches in jet study, see, e.g.,~\cite{Soper:2011cr,Ellis:2012sn,Ellis:2014eya,Mackey:2015hwa,Andreassen:2018apy}).  

Fourthly, apply the tool of persistent homology for the event-level data analysis at colliders. In this case the constituents in each event will be all projected to the detector sphere or cylinder. The persistent Betti numbers are expected to be applied to characterize topological structure of the whole event. One analogue with a reversed process is the generalization of the event shape N-jettiness~\cite{Stewart:2010tn} to the jet shape N-subjettiness~\cite{Thaler:2010tr}.  Notably, this topological observable scheme is generically insusceptible to the boost of collision events along beam direction, because of its topological nature, and hence can be well-applied to both hadron and lepton colliders. This application can be also extended to the resonance search, where the color of new particles may result in distinguishable topological structures in their decay products. The persistent homology thus may serve as a new type of color-flow observables~\cite{Gallicchio:2010sw}.   

Finally, develop more complete jet morphology by properly including geometric elements in this topological observable scheme. Recall, in integral geometry the shape of a $D$-dimensional geometric object is characterized with $D+1$ quantifiers, named Minkowski functionals~\cite{Minkowski1903}. If this object is smooth and closed, one of these quantifiers is reduced to Euler-characteristic, via Gauss-Bonnet theorem, an alternating sum of Betti numbers (see Eq.(\ref{eq:poincare})), while the other ones are purely geometric. This indicates that topological variants are not complete in characterizing jet structure and more complete jet morphology could be developed by incorporating geometric Minkowski functionals. Given the role of Minkowski functionals in analyzing morphology of the Cosmic Microwave Background (CMB) map~\cite{Mecke:1994ax,Winitzki:1997jj,Schmalzing:1997uc}, doing so will also enrich the dictionary between the collider observable scheme and the CMB one which was built recently~\cite{Li:2020vav}. Some of the topics in this wishlist are being investigated by this collaboration~\cite{wishlist}. 

\section*{Acknowledgements}
This research was supported partly by the General Research Fund (GRF) under Grant No 16305219 and partly by the Area of Excellence under the Grant No AoE/P-404/18-3. Both grants were issued by the Research Grants Council of Hong Kong S.A.R. This manuscript has been authored by Fermi Research Alliance, LLC, under Contract No. DE-AC02-07CH11359 with the U.S. Department of Energy, Office of Science, Office of High Energy Physics.

\bibliography{reference}
\end{document}